\def\Isimg{\raisebox{-.8ex}{$\stackrel{\textstyle >}{\sim}$}}

\def\Totem{\protect{\it {\sc Totem\/}}}

\def\bold#1{\setbox0=\hbox{$#1$}%
     \kern-.025em\copy0\kern-\wd0
     \kern.05em\copy0\kern-\wd0
     \kern-.025em\raise.0433em\box0 }
\def\slash#1{\setbox0=\hbox{$#1$}#1\hskip-\wd0\dimen0=5pt\advance
       \dimen0 by-\ht0\advance\dimen0 by\dp0\lower0.5\dimen0\hbox
         to\wd0{\hss\sl/\/\hss}}

\documentstyle[12pt,epsfig]{article}
\input{psfig.sty}

\newlength{\dinwidth}
\newlength{\dinmargin}
\newcommand{\resection}[1]{\setcounter{equation}{0}\section{#1}}

\begin{document}
\def\lq{\left [}
\def\rq{\right ]}
\def\LL{{\cal L}}
\def\VV{{\cal V}}
\def\AA{{\cal A}}

\newcommand{\be}{\begin{equation}}
\newcommand{\ee}{\end{equation}}
\newcommand{\bea}{\begin{eqnarray}}
\newcommand{\eea}{\end{eqnarray}}
\newcommand{\nn}{\nonumber}
\newcommand{\dd}{\displaystyle}

\thispagestyle{empty}
\vspace*{4cm}
\begin{center}
  \begin{Large}
  \begin{bf} 
    ROLE OF NEURAL NETWORKS IN THE SEARCH\\
    OF THE HIGGS BOSON AT LHC\\
  \end{bf}
  \end{Large}
  \vspace{1cm}
  \begin{large}
T. Maggipinto$^{a}$,G. Nardulli$^{a,b}$\\ 
  \end{large}
{\it $^{a}$ Dipartimento di Fisica, Universit\'a
di Bari, Italy}\\
{\it $^{b}$ Istituto Nazionale di Fisica Nucleare, Sez. di Bari, Italy}\\
  \vspace{8mm}
  \begin{large}
S. Dusini$^{c,d}$, F. Ferrari$^{c,f}$, I. Lazzizzera$^{c,d}$,
  A. Sidoti$^{c,d}$,\\
A. Sartori$^{d,e}$, G.P. Tecchiolli$^{d,e}$\\
  \end{large}
{\it $^{c}$ Dipartimento di Fisica, Universit\'a di Trento, Italy}\\
{\it $^{d}$ Istituto Nazionale di Fisica Nucleare, Gruppo Coll. di Trento, 
Sez. di Padova, Italy}\\
{\it $^{e}$ Istituto per la Ricerca Scientifica e Tecnologica, Trento,
Italy}\\
{\it $^{f}$  LPTHE, Universit\`es Paris VI - Paris VII, Paris, France }\\
  \vspace{5mm}
\end{center}
  \vspace{2cm}
\begin{center}
BARI-TH/268-97 \\
May 1997\\
\end{center}
\vspace{1.3cm}
\begin{quotation}
\begin{center}
  \begin{Large}
  \begin{bf}
  ABSTRACT
  \end{bf}
  \end{Large}
  \end{center}
  \end{quotation}
\begin{quotation}
\noindent
We show that neural network classifiers can be helpful to discriminate Higgs
production from background at LHC in the Higgs mass range $M_H\sim 200$GeV.
We employ a common feed-forward neural network trained by the
backpropagation algorithm for off-line analysis and the neural chip \Totem{}, 
trained by the Reactive Tabu Search algorithm, which could be used for on-line
analysis.
\end{quotation}
\newpage
\setcounter{page}{1}

\resection{Introduction}  

The main purpose of the future Large Hadron Collider (LHC) at CERN
is the search for the Higgs boson $H$, the only particle predicted by 
the Standard Model of the electroweak interactions that has not been 
discovered yet. The discovery of the Higgs particle would be of paramount
importance in confirming the peculiar feature of the electroweak vacuum 
embodied in the spontaneous breaking of the $SU(2)_L \times U(1)$ 
electroweak symmetry; on the other hand, its absence from the physical
spectrum would certainly pave the way for exciting new physics,
be it in the form of supersymmetry, or theories
with a strongly interacting Higgs sector \cite{cas} or something else.

The big theoretical and experimental effort that will be provided in
the next few years is strongly motivated by the relevance
of the stake, but also  because, as the Standard Model predicts and
detailed studies have confirmed \cite{lhc}, the signal, i.e. events
characterized by the production of the Higgs boson, will be
overwhelmed by background events, with multi-hadron production induced
by strong interactions of quark and gluons. For this reason, a crucial
step in the implementation of the LHC programme will be provided by
data analysis, which will be asked to disentangle Higgs events from huge
background. It is not our aim to review here the actual
experimental set-up of the LHC experiments, nor to examine the
performances of LHC detectors: the purpose of this letter is simply to
suggest that part of the task to extract the signal from the noise
could be supplied by Artificial Neural Networks (ANN)\footnote{for a
general introduction to ANN see \cite{hertz}.}

The role of ANN in high energy physics experiments has been stressed in a 
number of papers and we refer the interested reader to the existing
literature 
 \cite{lund1,alii,marchesini,defelice,brugnola,bortolotto,odorico}; 
in general, when compared with traditional methods of statistical
discrimination, they offer the advantage of possible on-line
implementation and often better results in terms of purity and
efficiency \cite{Den}. These latter features have been observed already in
preliminary studies on the use of ANN for Higgs search at LHC \cite{defelice}.
 The present analysis differs from these studies for two reasons:
\par\noindent
i) we adopt a more appropriate choice of the input variables;
\par\noindent
ii) we make a comparison between two implementations of ANN in the
feed-forward architecture; namely, a simulated neural network trained by
the usual backpropagation algorithm \cite{rum} is compared to a
hardware realization of a low-precision-weight Multi-Layer Perceptron
(MLP), the neurochip \Totem{} \cite{Pisa}, whose training-by-example task is
accomplished by a derivative free {\it combinatorial optimization}
algorithm called {\it Reactive Tabu Search}
(RTS) \cite{Bat-Tec-orsa,BatTec-ieee}.

A detailed presentation of the two NN will be given in Sections 2 and 3. 
Here we conclude this introduction by stressing the limits and some of the  
general features of our analysis.

First of all we do not investigate the whole expected Higgs mass ($M_H$)
range. From LEP data a lower bound for the Higgs mass is known:  $M_H\
\Isimg $ 60 GeV \cite{lephiggs} . Theoretical arguments based on 
unitarity or on the applicability of the perturbation theory indicate
that $M_H$ should not be larger than $\simeq 800$ GeV \cite{lus}; as for the
analysis based on
studies of radiative corrections, they appear still inconclusive, due to 
the weak dependence of these effects on $M_H$.
Since we mainly wish to present some case studies, rather than 
to make an extensive review, we limit our analysis to the mass values of
$M_H=150$ and  $200$ GeV. In a range around these Higgs mass values the
preferred decay channel, as indicated by previous studies 
\cite{lhc,defelice}, is the following one:
\be 
p ~p \to H~ X \to 4 \mu ~ X~ .\label{1} 
\ee
For rather larger Higgs masses ($\ge 400$GeV) the events (\ref{1})
would be clearly distinguishable from the peak in the four-muon
invariant mass.  In our case, however, the signal should be overwhelmed by
two main sources of background, namely the $t{\bar t}$ production:
\be
p~ p \to t~ {\bar t}~ X \to \mu^+\mu^-\mu^+\mu^- X^\prime, \label{2}
\ee
with the 4 muons arising from semileptonic decays of the top and antitop, and
the $Z b {\bar b}$ production:
\be
p~ p \to Z ~ b ~{\bar b}~ X\to \mu^+\mu^-\mu^+\mu^- X^\prime, \label{3}
\ee
with a muon pair arising from $Z^0$ decay and the other one from
semileptonic $b$ and ${\bar b}$ decays. It should be observed that,
due to the actual value of the top quark mass ($M_t=175\pm12$ GeV
\cite{mtop}), the processes (\ref{2}) and (\ref{3}) have comparable 
cross sections; for their calculation we rely in this paper on
the Pythia 
{\it Montecarlo} code \cite{pyt}.  At the LHC energy 
(${\sqrt s}=14$ TeV) one has\footnote{We notice that in
some previous studies, performed before the discovery of the top
quark, smaller values of $M_t$, e.g. $130$ GeV, were in general adopted,
which resulted in a higher cross section for the process (\ref{2}) and a
negligible one, in comparison, for the process (\ref{3}).}:
\bea
\sigma(pp\to t {\bar t} X \to \mu^+\mu^-\mu^+\mu^- X^\prime)&=& 
7.8\times 10^{-9} mb~, \label{4}\\ 
\sigma(pp\to Z^0 b {\bar b}X \to \mu^+\mu^-\mu^+\mu^- X^\prime)&=& 
6.0\times 10^{-9} mb~. \label{5}
\eea
These figures should be compared to the computed cross section for
Higgs production and subsequent decay into four muons: 
\be
\sigma(pp\to H ~X\to Z Z^\star~X  \to \mu^+\mu^-\mu^+\mu^- X)= 
1.2\times 10^{-12}mb \label{6}
\ee
for $M_H=150$ GeV;
\be
\sigma(pp\to H~X\to ZZ~X\to \mu^+\mu^-\mu^+\mu^- X)= 
2.8\times 10^{-12}mb \label{7}
\ee
 for $M_H=200$ GeV.  The main difference between the two cases is that 
for $M_H=200$ GeV
the two $Z$s are real, while for $M_H=150$ GeV only one is real,
the other being virtual. As a consequence, in the latter case
the constraint $M_{\mu \mu}=M_Z$ does not hold for one of the 
$\mu^+\mu^-$ pairs.

For the use of ANNs in high energy physics a crucial point is the
choice of sensible physical observables. On the base of previous studies 
\cite{lhc, defelice} the four final muons are ordered according to their 
energy, and the following 10
variables $X_1,...,X_{10}$ are introduced:

\begin{description}
\item[$X_1 - X_4$]: the transverse momentum of the four muons. 
The distributions of these variables for background events, as simulated
by the Pythia Montecarlo, show a maximum close to zero for those muons
coming from quark fragmentation, while the 
signal distributions show a peak around 25 -- 50 GeV; a similar distribution
is found for the
two muons deriving from $Z$ decay in the $Zb{\bar b}$ background events; 
\par\noindent
\item[$X_5 - X_8$]: the  invariant masses of the four different
$\mu^+\mu^-$ pairs.  For $M_Z < M_H < 2 M_Z$, two of the distributions
for the signal events show a peak around the $Z^0$ mass, that is absent for
the background. The peaks arise from muons coming from the real $Z^0$ decay;
they are two since the ordering based on the energy mixes in part the
events from the two $Z^0$. For $M_H \ge 2 M_Z$, of course, all the four
distributions exhibit the $Z^0$ mass peak;
\par\noindent
\item[$X_9$]: the four muons invariant mass;
\item[$X_{10}$]: the hadron multiplicity related to hard jets.
\end{description}
 
A comment on the variable $X_{10}$ is in order. We expect that hadrons 
generated by hard parton scattering are more 
copiously produced by the process (\ref{3}) and especially
(\ref{2}) as compared to (\ref{1}). However such a peculiar feature of 
the events (\ref{1}) is hidden in the huge number (typically several 
hundreds at the LHC energy) of hadrons produced by the hadronization 
of the two beam jets. The remnants of the two beams disintegration could 
be eliminated in the LHC experimental conditions by appropriate cuts 
in the angular variables, but in our simulations we choose to pre-process 
the data by the so called $k_{\perp}$ clustering algorithm
\cite{kt, catani}. This algorithm consists in general of
two steps. In the first step one compares
\be
d_{ij}\,=\,2\min\{E_{Ti}^2,E_{Tj}^2\}
\sqrt{(\eta_i-\eta_j)^2+(\phi_i-\phi_j)^2} \label{8}
\ee
and
\be
d_{iB}\,=\,E_{Ti}^2 \; ,\label{9}
\ee
where $E_{Ti}$ is the transverse energy of the $i$-th particle with 
respect to the beam direction, $\eta_i$ is its pseudorapidity and $\phi_i$
is the azimuth angle with respect to the beam axis: a final state
particle $i$ is attributed to the beam remnants (beam jet) if
$d_{iB}$ is smaller than $d_{ij}$, otherwise it is attributes to a hard jet.
In the second step, which is not of interest here, the particles belonging
to hard jets are divided into different clusters\footnote{For another
application of $k_\bot$ algorithm in the context of ANN studies of high
energy physics experiments see \cite{guido}.}. After the application of
the $k_{\perp}$ algorithm and the removal of the hadrons belonging to the
beam jets, the remaining hadron multiplicity is called by us $X_{10}$.
The relevance of the variable $X_{10}$ can be seen from Fig. 1, where we
compare its distribution relative to the processes (\ref{1}), (\ref{2}) and 
(\ref{3}).

Having defined the input variables, we now discuss the analyses
performed on the Montecarlo data using the two neural networks. 

\resection{Analysis by simulated ANN trained through the backpropagation
algorithm }

First we discuss the results obtained using a simulated net in the most
common architecture adopted in high energy applications, i.e. the {\it
feed-forward} MLP, trained though the "classical" backpropagation
algorithm.  The net is composed of one input layer with $10$ neurons $X_j$,
one hidden layer with $21$ neurons $z_j$ and one output unit $y$.

The physical observables introduced above, once normalized to the interval
$[-1, 1]$, become the inputs $X_j \; (j=1,...10)$  of the NN
classifier. Each pattern-event $p$ consists of the array $X_j$ of the
input variables (features) and the value $y$ of the output neuron
($y=1$ for the signal, i.e. the Higgs production, and $y=0$ for the
background).  The patterns have been divided into two sets, the
training set, used by the network to learn, and the testing set, used
to evaluate subsequently its performance.

As already mentioned, our simulations have been obtained by the Pythia
Montecarlo Code\cite{pyt}. We have treated the case of two possible
values for the mass of the Higgs particle: one below $2 M_Z$
i.e. $150$ GeV, and one just above, i.e. $200$ GeV. For each of these
mass values the training data set consisted of $N=2000$ signal events,
2000 $t\overline t$ and 2000 $Zb\overline b$ events, while the testing
data set consisted of 2000 $pp \to H X \to 4 \mu X$ signal events, $5.6
\times 10^6$ $t\bar t$ and $4.2 \times 10^6$ $Zb \bar b$ background
events for the case $M_{H}=200$ GeV. 
The data in the training sets were all different from those in
the testing sets.

As usual, the performance of the network has been evaluated by introducing
two variables: the purity ($P$) and the efficiency ($\eta$) defined as
follows:
\be
P \, =\, \frac{N^a_H}{N^a_H + N^a_B}
\ee
and 
\be
\eta \, = \, \frac{N^a_H}{N_H}
\ee
where $N_H$ is the total number of Higgs events in the testing sample, $N^a_H$
is the total number of the accepted (i.e. correctly identified) Higgs
events and $N^a_B$ is the total number of the accepted background events,
i.e. events that are incorrectly identified as Higgs events.

One can increase the purity decreasing the efficiency by introducing a
threshold parameter $l\ \epsilon\ [0,\ 1]$ as follows. The range of values
of the output neuron $y^{(p)}$ in the testing phase is divided into
the subintervals: $I_1 = [0,\ 1-l]$ and $I_2 = ]1-l,\ 1]$, so that if 
$y^{(p)}\ \epsilon\ I_1$ (respectively  $y^{(p)}\ \epsilon\ I_2$) the event is
classified as background  (respectively: signal).

Our results are reported in Fig. 2 (dots); it shows that in the case of 
$M_H \, = \, 200$ GeV  one can reach appreciable values of purity. The
situation is less favorable in the case of $M_H \, = \, 150$ GeV, when,
due to the virtuality of one of the two $Z$, the reduction of efficiency
is relevant.

\resection{Analysis using the RTS training algorithm as implemented on the
neurochip  \Totem{}}

One of the purposes of the present work is to contribute to clarify the
possibility of using neural network classifiers in time-critical
operations, like the fast triggering required in some high energy
experiments, without loosing high quality performances.
The neurochip \Totem{}, has been conceived to implement
Multi-Layer Perceptrons in the {\it feed-forward} architecture on the basis
of a simple and fast computational structure \cite{Pisa}. This is
achieved escaping the necessity of derivative calculations, turning
the task of training-by-examples into a {\it combinatorial
optimization} problem, whose solution is searched then by means of the
{\it Reactive Tabu Search} method \cite{Bat-Tec-orsa, BatTec-ieee}.
Differently from the derivative-based backpropagation algorithms, RTS
thus allows simple and low precision computation, using only up to 8
bits for the synaptic weights and 16 bits to represent the feature
parameters\footnote{For a comparison in performance between \Totem{}
and backpropagation based neurochips see for instance
\cite{TOT_et_al}}: this is indeed the basis of the simple and fast
computational structure said above.
\Totem{} can be set to different {\it feed-forward} MLP architectures and
for the present work it has been given exactly the same 10-21-1
architecture as the network described in the preceding section. We
have used the same simulation Pythia Montecarlo data as before, as well as the
same overall procedure (not the algorithm, evidently) for training and
testing. A remarkable difference is that now we represent the physical
observables by five decimal digit integers, against  the double
precision floating point representation needed for the
backpropagation NN. The {\it truth} value for a Higgs event was fixed
to 8192 and to 0 for a background events, while the threshold
parameter controlling the purity level was varied by steps of unity
between the two {\it truth} values.
The results in terms of purity  and efficiency are collected in Fig. 2
for a case of 8-bit synaptic weights and 200 GeV of Higgs mass (the data are 
represented by stars).

\resection{Conclusions}

Neural networks have a clear advantage over traditional statistical methods, 
since they
can support a high degree of parallelism and could be used for on-line
analysis of the experimental data. Therefore their use in the future
LHC experiments should be seriously considered and thoroughly investigated.
We have contributed to this analysis by considering two different nets. 
The first one is a simulated ANN trained by the backpropagation
algorithm.  The second one is a hardware implementation of a fast  NN,
the neurochip \Totem{}.

Our results show that NN can be helpful in the discrimination of
background events from the signal in the Higgs search at the future
Large Hadron Collider to be built at CERN. We have proved this by
considering one particular Higgs decay channel ($H \to 4 \mu$)
in the mass range $M_H\ \epsilon\ (150-200)$ GeV and including the
most relevant backgrounds: $t {\bar t}$ and $Z b{\bar b}$.
For both the neural nets, the case $M_H\simeq 200$ GeV is more favourable,
and acceptable values of purity and efficiency can be obtained; in particular
the neural chip \Totem{} produces in general better performances and, in view
of its possible on-line implementation, should be seriously considered, in
our opinion, as a tool for the analyses to be performed at the future
Large Hadron Collider at Cern.

{\bf Acknowledgments.} We wish to thank G. Marchesini for most useful
comments and P. De Felice and G. Pasquariello for their collaboration at an
early stage of this work.

\newpage
\centerline
{\bf FIGURES.}
\begin{figure}[h]
\vspace{8cm}
\includegraphics{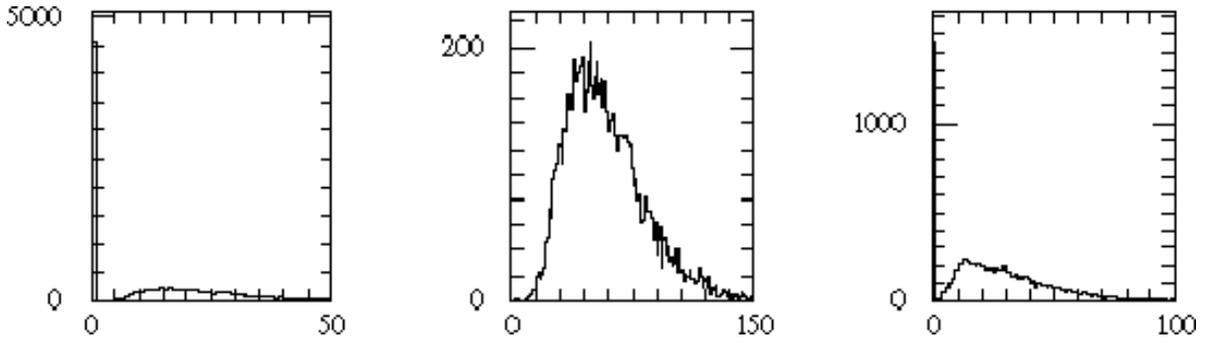}
\vspace{-2cm}
\caption{$X_{10}$ distribution for, from the left, Higgs signal,
$t \bar t$ and $Zb \bar b$ background.}
\end{figure}

\begin{figure}[h]
\vspace{14cm}
\includegraphics{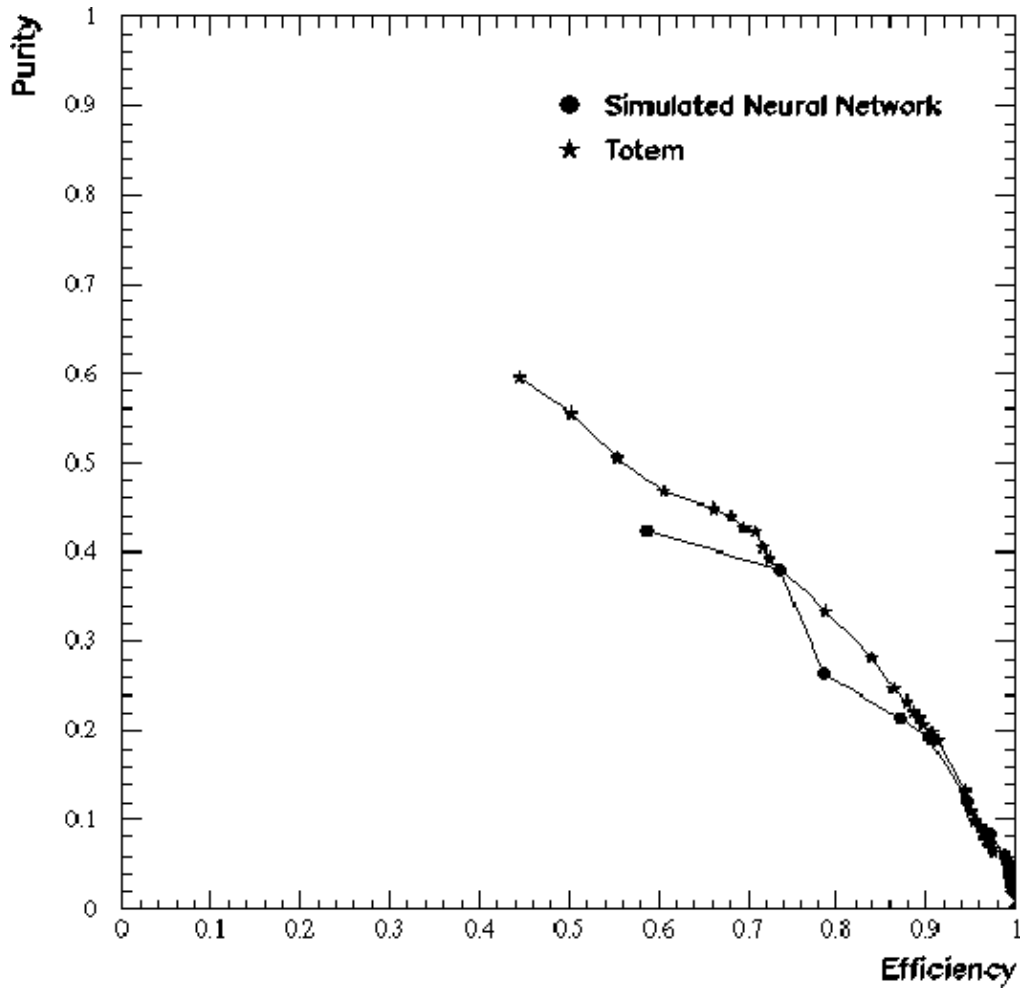}
\caption{The purity $P$ versus the Higgs efficiency
$\eta$ for two different NN in the case $M_H=200$ GeV.}
\end{figure}
\end{document}